\documentclass[lettersize,journal]{IEEEtran}
\usepackage{graphicx}
\usepackage{cite}
\usepackage{picinpar}
\usepackage{amsmath}
\usepackage{url}
 \usepackage[utf8]{inputenc}
\usepackage{soul}
\usepackage[ruled,vlined]{algorithm2e}
 \usepackage{mathrsfs} % for \mathscr
\usepackage{multirow}
 \usepackage{pifont}
\usepackage{mathtools}
 \usepackage{color}
 \usepackage{easyReview}
\usepackage{etoolbox} % Ensure compatibility

\usepackage{fix-cm} % Ensures compatibility with font size commands
\usepackage[dvipsnames]{xcolor}
 \usepackage{alltt}

\usepackage[mathscr]{euscript}
\usepackage{siunitx}
 \usepackage{epstopdf}
 \usepackage{pbox}
 \usepackage{amssymb}

\usepackage{comment}
 \usepackage{balance}

\usepackage{algpseudocode}

\usepackage[export]{adjustbox}
\usepackage{float}
 \usepackage{pdfpages}

 \usepackage{bm}

\newtheorem{theorem}{Theorem}
\newtheorem{remark}{Remark}

\newtheorem{lemma}{Lemma}
\newtheorem{proposition}{Proposition}

\begin{document}
%\begin{frontmatter}

\title{A Novel Online Pseudospectral Method for Approximation of Nonlinear Systems Dynamics} %\thanksref{footnoteinfo}}

\author{Arian Yousefian$^{1}$, Avimanyu Sahoo$^{1}$, and Vignesh Narayanan$^{2}$% <-this 
\thanks{
This work is partially supported by NSF \#2327409, \#2337998, and \#2337999, and NSWC-PC \#N00174-23-1-0006. 

$^{1}$Arian Yousefian and Avimanyu Sahoo are with the Electrical and Computer Engineering Department at the University of Alabama in Huntsville, Huntsville, AL (Emails: {\tt\small ay0024@uah.edu}, and {\tt\small avimanyu.sahoo@uah.edu}), and $^{2}$Vignesh Narayanan is with the AI Institute, University of South Carolina (Email: {\tt\small vignar@sc.edu}).
}
}

\maketitle
\begin{abstract}
This note presents an online pseudospectral method for system identification using Chebyshev polynomial basis under aperiodic sampling.  The system dynamics are approximated piecewise by introducing a sliding time window. The number of sampling instants (Chebyshev nodes) within each sliding window is selected dynamically based on a proposed node-selection criterion that guarantees desired approximation accuracy. The system states are measured at these aperiodic instants and used to estimate the coefficients of the basis polynomials using least squares. An adaptive state estimator is also proposed to reconstruct the continuous states using the approximated dynamics. The boundedness of the parameter and state estimation errors is proven analytically and validated numerically.
\end{abstract}


%\end{frontmatter}

%%================================================
\section{Introduction}
System identification (ID) is fundamental to engineering systems because of unavailability of precise models due to complex and uncertain dynamics. It facilitates data-driven modeling essential for analysis, control, and decision making in applications such as robotics, aerospace vehicles, and power networks \cite{Nelless2020}. While linear ID is well studied, nonlinear ID remains challenging, especially online, where models are updated in real time \cite{ljung2010perspectives}.  
 Recent efforts on system ID span classical parametric schemes, spectral methods \cite{jin2024extended} and pseudospectral (PS) methods \cite{pirastehzad2021successive}. Spectral techniques rely on eigenfunction decompositions \cite{mezic2005spectral}, whereas PS methods employ orthogonal polynomial collocation \cite{constantine2012sparse}. Classical orthogonal polynomial families (Legendre, Chebyshev, Hermite, Laguerre) underpin these methods (see, for example, \cite{trefethen2019approximation}).

The PS methods, particularly those employing Chebyshev polynomials, are attractive due to minimax uniform approximation, node clustering that mitigates Runge oscillations, and efficient coefficient computation \cite{mason2002chebyshev,rivlin2020chebyshev,trefethen2019approximation}. Most PS ID schemes, however, assume offline data over a fixed domain \cite{ma2010efficient}. In parallel, neural networks (NNs) \cite{lewis2020neural} and event-based NN training approaches \cite{sahoo2015adaptive} for online approximation have been developed. These approaches typically require higher sampling during learning \cite{sahoo2015adaptive} and the approximation accuracy depends on the number of neurons \cite{lewis2020neural}.

In our previous work \cite{yousefian2024aperiodic}, an online Chebyshev-based PS scheme with a sliding  window was proposed where  sampling time instants were chosen at Chebyshev nodes, and state measurements at those instants were used to construct the basis. Because the state measurements were not the roots of the Chebyshev polynomial (Chebyshev nodes) of the basis, uniform approximation-error bounds were not guaranteed. To address this gap, this note introduces an online Chebyshev PS approach for ID that guarantees a uniform approximation with a desired accuracy under reduced, aperiodic sampling. 

Instead of using states, the Chebyshev basis polynomial is constructed as a function of time over arbitrary intervals. The number of Chebyshev nodes (which determines the basis polynomial order and sampling points) within each sliding time window is selected adaptively to meet the desired accuracy. The coefficients of the basis function are estimated by least squares regression from the measured states and their derivatives at Chebyshev time nodes within each window, yielding a piecewise approximation of the system dynamics.  The identified dynamics are then used to design an adaptive state estimator. To ensure continuity across sliding-windows, the approximation coefficients are recomputed so that the estimated dynamics match at each window transition point. 

Furthermore, a detailed errors analysis and adaptive selection of the number of Chebyshev nodes, establishing a criterion for approximation order and sampling are presented. The boundedness of both the parameter and state estimation errors is proven analytically and, further, substantiated through simulation using a benchmark nonlinear oscillator system. This work builds on the preliminary work presented in \cite{yousefian2025novel}. The main contributions of this letter, beyond the preliminary work in \cite{yousefian2025novel}, are: i) an online Chebyshev PS ID method; ii) an adaptive node-selection strategy with a theoretical guarantee on the desired approximation accuracy; iii) an adaptive PS state estimator; and iv) the convergence analysis establishing boundedness parameter and state estimation errors.

This note is organized as follows. Section \ref{back_ground} reviews offline PS methods and formulates the problem. Section \ref{sec:proposed_ps_scheme} introduces the online PS framework, followed by Section \ref{sec:state_estimation}, which discusses adaptive state estimation and node selection. Section \ref{simulation_results} presents the simulation results to validate the analytical claims, followed by Conclusions in Section \ref{sec:conclusion}.

\textbf{Notations.} $\mathbb{R}^n$ is the $n$-dimensional Euclidean space; $\mathbb{R}^{n\times m}$ the set of $n\times m$ real matrices. 
$\|\cdot\|$ denotes the Euclidean norm. The superscript $(\cdot)^\top$ denotes transpose. For a matrix A, \text{Sym(A)} denotes $A+A^\top$. The ceiling and floor operators and Kronecker product are denoted by $\lceil\cdot\rceil$,  $\lfloor\cdot\rfloor$, and   $\otimes$, respectively. If \(x(t):\mathbb{R}\to\mathbb{R}^{n}\) is essentially bounded, then it is in the set $\mathcal{L}_\infty^{\,n} :=  \{ x(t) | \|x\|_{\infty} \triangleq \sup_{t\ge 0}\|x(t)\| < \infty   \}$, the set of all real-valued functions defined on the interval \([a,b]\) that are \(N\)-times continuously differentiable is denoted by \(C^N([a,b])\).

%================================================
\section{Background on Chebyshev Polynomial-based Approximation and Problem Statement}\label{back_ground}

This section briefly reviews offline PS approximation with Chebyshev polynomial bases and formulates the problem. 

\subsection{Chebyshev Approximation on the Standard Interval}
Let $g(x) \in C^{ N+1}[-1, 1] $ be a scalar function. Using a first-kind Chebyshev PS basis $T_i(x)$, $g(x)$ admits the expansion
\begin{equation}\label{eq: chebyshev_apprxox_Fff}
    g(x) = \sum\nolimits_{i= 0}^\infty c_i^* T_i(x)  = \sum\nolimits_{i=0}^{N} c_i^* T_i(x) + \epsilon(x),
\end{equation}
where $\{c_i^*\}_{i=0}^{N}$ are the unknown coefficients and $\epsilon(x)$ is the truncation error and  \(\{T_i(x)\}_{i=0}^{N}\) on $[-1,1]$ are defined recursively as
\begin{equation}\label{eq:chebyshev_poly_iterative}
    \begin{cases} 
        T_0(x) = 1, \quad T_1(x) = x, & \\ 
        T_i(x) = 2xT_{i-1}(x) - T_{i-2}(x), & i = 2, 3, \ldots, N.
    \end{cases}
\end{equation}
The roots of $T_i(x)$, called the \textit{Chebyshev nodes}, are widely used in polynomial interpolation. For $T_N(x)$, the nodes are
\begin{equation}\label{EQPAR1}
    x_k = \cos\left((k-0.5)\pi/{N}\right), \quad k = 1, 2, \ldots, N.
\end{equation}
The nodes are non-uniformly spaced and denser near the interval endpoints  \cite{rivlin2020chebyshev}.

In practice, the domain often differs from $[-1,1]$. Therefore, an affine map can be used to transform $x$ in $[-1, ~1]$ to $[a, b]$ as
\begin{equation}\label{shifted node}
    x^S= \frac{1}{2}\left((a+b) + (b-a)x\right),\qquad a\le x^S\le b.
\end{equation}
Thus, the Chebyshev basis on $[a,b]$ can be obtained by applying the change of variable in \eqref{shifted node}, which map it back to $[-1 ~ 1]$, as
\begin{equation}\label{shifftedpoly}
T_i^S(x^S)=T_i\!\left((2x^S-(a+b))/({b-a})\right).
\end{equation}
Similarly, the nodes on $[a,b]$ can be computed as
\begin{equation}\label{shifted node11}
    x^S_k= \frac{1}{2}\left((a+b) + (b-a)x_k\right),\quad k=1,2,\ldots,N.
\end{equation}
For brevity, hereafter, we use $x$ for the shifted variable $x^S$ and use the superscript $S$ to indicate shifted polynomials.

\subsection{Chebyshev Expansion on a Shifted Interval}
The function $g(x)$ on $[a,b]$ admits the shifted expansion  $g(x)  = \sum\nolimits_{i=0}^{N} c_i^{S^*}{T_i^S(x)} + \epsilon(x)$, where \(\{c_i^{S^*}\}_{i=0}^{N}\) are the unknown coefficients in the shifted Chebyshev basis, and the PS approximation is given by
\begin{equation}\label{eq: chebyshev_apprxox_F_app}
    \hat{g}(x)  = \sum\nolimits_{i=0}^{N} c^S_i T_i^S(x),
\end{equation}
where \(c_i^{S}\) denotes the estimates of \(c_i^{S^*}.\)

To estimate \(c_i^{S}\), one can sample \(y_k = g(x_k)\) at \(m \ge N+1\) shifted Chebyshev nodes \(x_k \in [a,b]\) and employ least squares, given by \cite{mason2002chebyshev}
\begin{equation}\label{eq:ols_alpha}
c^S = (X^\top X)^{-1} X^\top Y,
\end{equation}
where ${{c^S}}={{\left[ \begin{matrix}
   c^S_{0} & c^S_{1} & \ldots  & c^S_{N}  \\
\end{matrix} \right]}^{\top}}$, and 
$$X =
\begin{pmatrix}
T_0^S(x_1) & T^S_1(x_1) & \dots & T^S_N(x_1) \\
T_0^S(x_2) & T^S_1(x_2) & \dots & T^S_N(x_2) \\
\vdots & \vdots & \ddots & \vdots \\
T_0^S(x_m) & T^S_1(x_m) & \dots & T^S_N(x_m)
\end{pmatrix}, \,\,\,
Y =
\begin{pmatrix}
y_1 \\ y_2 \\ \vdots \\ y_m
\end{pmatrix}.$$

From an interpolation error perspective, for each \(x \in [a,b]\) there exists a point \(\xi_x \in [a,b]\) such that \cite{stewart1996afternotes}
\begin{equation}\label{eq:general_error}
    g(x) - \hat{g}(x) = \frac{g^{(N+1)}(\xi_x)}{(N+1)!} \prod \nolimits_{k=1}^{N+1}(x - x_k).
\end{equation}

Since Chebyshev nodes minimize the maximum interpolation error, with these nodes, the error is bounded by
\begin{equation}\label{eq:chebyshev_error}
    \max_{x \in [a,b]} \big|g(x) - \hat{g}(x)\big| \leq \frac{2D^{(N+1)}}{(N+1)!} \left(\frac{b-a}{4}\right)^{N+1},
\end{equation}
where $D^{(N+1)}=\underset{\xi_x \in [a,b]}{\mathop{\max }}\,\big|g^{(N+1)}(\xi_x)\big|$.

The PS methods are typically leveraged for offline learning using pre-collected data with known $[a,b]$ and order $N$. However, online PS approximation introduces several challenges and are detailed below.

%=================================================
\subsection{Problem Formulation}
Consider a continuous-time nonlinear system described by
\begin{equation}\label{Eq: dyanmicform_closed_loop1}
  \dot{x}(t)=F\bigl(x(t)\bigr),
\end{equation}
where \(x(t)\in\mathbb{R}^{N_{\mathscr{P}}}\) is the state vector and
\(F:\mathcal{D}\to\mathbb{R}^{N_{\mathscr{P}}}\) denotes
the unknown nonlinear dynamics. The domain \(\mathcal{D}\subset\mathbb{R}^{N_{\mathscr{P}}}\)
contains the origin, and \(F(0)=0\).
The function \(F\) is locally Lipschitz, and \(x(t)\in \mathcal{L}_\infty^{\,N_{\mathscr{P}}}\). The system in \eqref{Eq: dyanmicform_closed_loop1} can be considered as a closed-loop system or an autonomous system and the identification scheme can be used for prediction or control design.

The objective is to approximate \(F\big(x(t)\big)\) online via the PS method, using Chebyshev polynomials of the first-kind as basis functions, with aperiodically sampled data at Chebyshev nodes to ensure the desired accuracy. Equivalently, we seek an approximation $\hat{F}(x(t)) \in \mathcal{Q}_M$, the set of degree-$M$ polynomials, that minimizes the interpolation error
\begin{equation}\label{h_infty_approx_error}
\mathscr{E}\big(x(t)\big) = \min_{\hat{F}\big(x(t)\big) \in \mathcal{Q}_M} \left\| F\big(x(t)\big) - \hat{F}\big(x(t)\big) \right\|_{\infty}.
\end{equation} 
For \( x(t) \in \mathbb{R}^{{N}_{\mathscr{P}}} \), the dynamics  \eqref{Eq: dyanmicform_closed_loop1} 
can be approximated as
 \begin{equation}\label{n-dimensional_basis}
\hat{F}\big(x(t)\big) =\eta^{\top} \bigotimes_{j=1}^{{N_\mathscr{P}}} \mathbb{T}^S_{M}\big(x^j(t)\big),\,j = 1,2,\ldots,N_\mathscr{P},
 \end{equation}
for $x^j(t) \in [a^j, b^j]$, where \(\mathbb{T}^S_{M}\big(x^{j}(t)\big)
= \big[T_{0}^S\big(x^{j}(t)\big),\,T^S_{1}\big(x^{j}(t)\big),\,\ldots,\,T^S_{M}\big(x^{j}(t)\big)\big]^{\top}\)
is the shifted Chebyshev basis vector, \(\eta \in \mathbb{R}^{(M+1)^{N_{\mathscr{P}}}\times N_{\mathscr{P}}}\) is the estimated-coefficient matrix \cite{yousefian2024aperiodic}, and \(M\) is the node count.

Computing Chebyshev nodes using \eqref{shifted node11} requires prior knowledge of the interval $[a^{j},b^{j}]$ and the node count $M^{j}$ for each state $x^{j}(t)$. Since the dynamics are unknown, the bounds $a^{j},b^{j}$ and the required $M^{j}$ cannot be specified apriori, and the nodes cannot be precomputed. Further, to measure the states at the nodes, i.e., $\{x^{j}_{k}\}_{k=1}^{M^{j}+1}$,  requires the corresponding times $\{t^j_k\}_{k=1}^{M^j+1}$. Since there is no one-to-one map from state values to time, the sampling times cannot be determined for sampling and state measurement at nodes.

Therefore, for online Chebyshev PS approximation, the problem is fourfold: (i) to develop an online method that does not require sampling at state-space Chebyshev nodes; (ii) to determine suitable approximation intervals during operation to enable data collection; (iii) to select the polynomial degree or node count to meet a desired approximation accuracy; and (iv) to reconstruct the continuous state from aperiodic measurements. The next sections present a detailed solution.

%==========================================
\section{Proposed Online PS Framework}\label{sec:proposed_ps_scheme}
\begin{figure}[!ht]
    \centering
\includegraphics[width=\linewidth]{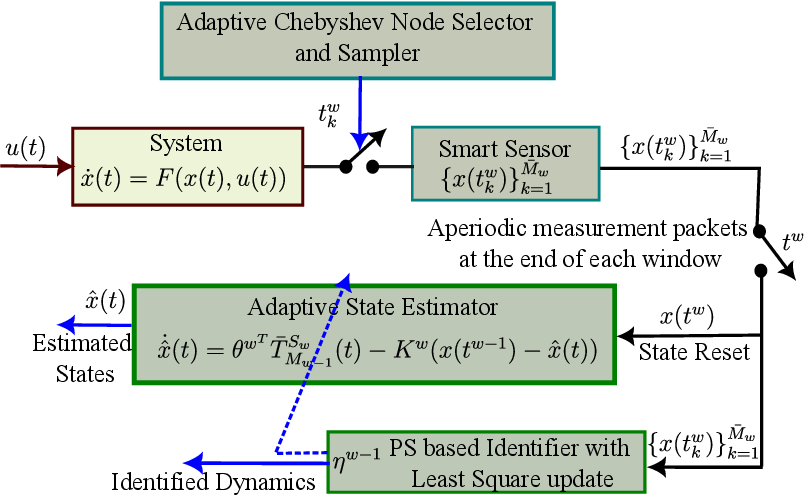}
    \caption{Schematic of proposed approach.}
    \label{fig:ps_block_diagram}
\end{figure}

In this section, an online approximation framework based on the PS method is proposed using a sliding time window and the minimum number of nodes to guarantee desired approximation accuracy, as illustrated in Fig.~\ref{fig:ps_block_diagram}. 

Within each window, a smart sensor samples the system states at time nodes determined by the adaptive selector, packages the data, and transmits it to the identifier at the end of each window. The identifier performs batch least squares estimation to obtain coefficients (parameters) of the basis function. These identified parameters are then used to construct the adaptive state estimator that can estimate the system states from aperiodic sampling for the next window. The number and location of nodes, which determine both the sampling instants and the polynomial approximation order, are updated adaptively to ensure the desired approximation accuracy while minimizing computational effort.

\subsection{Online PS Method for System Identification}
The dynamics in \eqref{Eq: dyanmicform_closed_loop1} are an implicit function of time. 
Each component $F^{j}(x(t))$, $j=1,\ldots,N_{\mathscr{P}}$, can be approximated by the first-kind Chebyshev polynomial basis with time as the argument \cite{ma2010efficient}, given by
\begin{equation}\label{eq: chebyshev_apprxox_Ff}
    F^{j}(t) = \sum\nolimits_{i=0}^{M} \eta_{ij}^{*}\,T_{i}(t) + \epsilon_{j}(t),
\end{equation}
where  
\(\eta_{ij}^{*}\in\mathbb{R}\) are the unknown coefficients,  
\(\epsilon_{j}(t)\in\mathbb{R}\) is the truncation error, and  $M$ is approximation order (node count). The approximation of the function can be expressed as
\begin{equation}\label{eq: chebyshev_apprxox_Ff1}
    \hat{F}^j(t) =  \sum\nolimits_{i=0}^{M} \eta_{ij} T_i(t),
\end{equation}
where $\eta_{ij} \in \mathbb{R}$ are the estimates of the target coefficients $\eta_{ij}^*$.

\begin{remark}
Constructing the Chebyshev basis as a function of time avoids using a high-dimensional state basis for $N_{\mathscr{P}}$-dimensional functions in \eqref{n-dimensional_basis}. This also allows online approximation and determining sampling points at time nodes. 
\end{remark}
 
To enable online approximation of the dynamics $F(t)$, define a sliding window $(t^{w-1},t^{w}]$ $w= 1,2, \cdots$, 
where $t^{w}$ denotes the transition between consecutive windows and 
$\tau^{w}=t^{w}-t^{w-1}$ is the window width.
 %In each window $w$, the approximation order $M_{w}$ is adaptively updated  until the error falls below a desired approximation error (see Section~\ref{node_selection}).  
Using \eqref{shifted node11},  the \( M_w + 1 \) time nodes (sampling instants) within \( (t^{w-1}, t^{w}] \) can be computed as
\begin{equation}\label{eq: time_nodes}
t^{w}_{k} = \frac{1}{2}\left( t^{w-1} + t^{w} \right) + \frac{1}{2} \tau^w \cos\left(\frac{k- 0.5}{M_w + 1} \pi\right),
\end{equation}
for \( k = 1, 2, \dots, M_w+1 \). 

\begin{remark}
 For ease of exposition, in this paper,  we considered constant window width, i.e., \( \tau^w = \tau\), where $\tau$ is a constant. A variable \( \tau^w \) can also be designed with the use of event-triggered or self-triggered control strategies \cite{sahoo2015neural}, where the sampling intervals are adaptively determined based on stability and performance criteria in a controlled system.
\end{remark}

With time nodes as in \eqref{eq: time_nodes}, the sensor samples $x(t_k^{w})$ at each Chebyshev node $t_k^{w}$ in window $w$ and also $x(t_k^{w}-\Delta t)$ with $\Delta t>0$ to estimate $\dot x(t)$ at the nodes. In addition, the state is measured at the start of each window to reset the identifier (discussed later in Sec. \ref{sec:state_estimation}). All samples from the $w$-th window are used  to estimate the coefficients %to as a single packet to the identifier, as shown in Fig. \ref{fig:ps_block_diagram} for coefficient estimation, and the coefficients are updated only 
at window transition instants.

\begin{remark}
For simplicity, state derivatives are approximated by a backward difference, which is known to amplify noise. To address this challenge, one can use adaptive or sliding mode state derivative estimators \cite{trefethen2019approximation,levant2003higher}.
\end{remark}

From \eqref{eq: chebyshev_apprxox_Ff}, the piecewise representation of the 
$j$-{th} system dynamics for  $j =1, 2, \cdots, N_p$ within $w$-th window is
\begin{equation}\label{eq: chebyshev_apprxox_Ff_w}
    F^j(t) = \sum\nolimits_{i=0}^{M_w} \eta_{ij}^{{w^*}} T_i^{S_w}(t) + \epsilon^w_{j}(t),\,\,\, t^{w-1}< t\le t^w,
\end{equation}
where $T_i^{S_w}(t)$ is the shifted Chebyshev polynomial on the window $t^{w-1}< t\le t^w$. 
Defining the basis vector as 
\begin{equation}\label{vector basisss}
\bar{T}_{{M}_{w}}^{{S}_{w}}(t) =
\left[ 
\begin{matrix}
   T_{0}^{{S}_{w}}(t) & T_{1}^{{S}_{w}}(t) & \cdots & T_{{M}_{w}}^{{S}_{w}}(t)
\end{matrix}
\right]^{\top},
\end{equation}
the augmented system dynamics $F^w(t) \in \mathbb{R}^{N_\mathscr{P}}$ in \eqref{eq: chebyshev_apprxox_Ff_w} can be expressed as
\begin{equation}\label{eqPNnnnnn1}
F^w(t)={{\eta }^{{w^*}^\top}}\bar{T}_{M_w}^{S_w}(t)+\epsilon ^{w},\quad t^{w-1} < t \le t^w,
\end{equation}
where  $\eta^{w^*}=\left[ \begin{matrix}
   \eta_{1}^{w^*} & \eta_{2}^{w^*} & \cdots  & \eta^{w^*}_{N_\mathscr{P}}  \\
\end{matrix} \right] \in  \mathbb{R}^{\left( {M}_{w}+1 \right) \times {N_\mathscr{P}}}$ with $\eta_{j}^{w^*} \in \mathbb{R}^{\left( {M}_{w}+1 \right)}$ is the parameter matrix  for the truncated Chebyshev polynomial. The approximation of \eqref{eqPNnnnnn1} on $w$-th window can be expressed as
\begin{equation}\label{eqPNnn1}
\hat{F}^w(t)={{\eta }^{{w}^\top}}\bar{T}_{M_w}^{S_w}(t), \quad t^{w-1} < t \le t^w,
\end{equation}
where $\eta^w\in\mathbb{R}^{\left( {M}_{w}+1 \right) \times {N_\mathscr{P}}}$ is the estimated  coefficients. 

Using the $w$-th window measurements at $\{t_k^{w}\}_{k=1}^{M_w+1}$, the least-squares estimator in \eqref{eq:ols_alpha} provides $\eta^{w}$ at $t^{w}$ as 
\begin{equation} \label{eqPN12}
\eta^{w} \!=\! \big({\bar{\mathbb{T}}}^{S_{w}}_{M_w}(t) {\bar{\mathbb{T}}}^{S_{w}^{^\top}}_{M_w}(t) +\! R_0^w \big)^{-1}\! 
\big(R^w_0 \eta_{0}^w \!+\! {\bar{\mathbb{T}}}^{S_{w}}_{M_w}(t) {\dot{X}}^{w}(t)\big),
\end{equation}
where $\bar{\mathbb{T}}_{M_w}^{S_w}(t)\in \mathbb{R}^{(M_w+1) \times (M_w+1)}$   and  $\dot{X}^{w}\in \mathbb{R}^{(M_{w}+1) \times N_\mathscr{P}}$  with 

$\bar{\mathbb{T}}_{M_w}^{S_w}(t) ={{\big[ \begin{matrix}
   \bar{T}_{M_w}^{S_w}(t_{1}^w) & \bar{T}_{M_w}^{S_w}(t_{2}^w)  \cdots   \bar{T}_{M_w}^{S_w}(t_{(M_{w}+1)}^w)  \\
\end{matrix} \big]}}  $ and $\dot{X}^{w} ={{\big[ \begin{matrix}
   \dot{x}(t_1^{w}) & \dot{x}(t_2^{w}) & \cdots  & \dot{x}(t^w_{{(M_{w}+1})})  \\
\end{matrix} \big]}^{\top}} $ for $t^{w-1} <t \le t^w$. Matrices $R_0^{w}$ and $\eta_0^{w}$ are user-defined to ensure the first-term inverse in \eqref{eqPN12} exists \cite{krstic1995nonlinear}.

\subsection{Minimum Nodes for Desired Approximation Accuracy}

The key advantage of piecewise approximation with a moving window is the ability to dynamically choose the degree of the polynomial $M_w$ to keep the approximation accuracy within a desired bound in each window. The following theorem provides the criteria for selecting the minimum node count using the maximum error per-window. Before introducing the node selection criteria, the following lemma is necessary.

\begin{lemma}\label{lm:exponential_bound_max_error}
Define $\mathscr{E}^w_{\max}=\max_{t^{w-1}<t\le t^{w}}\|F^w(t)-\hat{F}^w(t)\|$.
For the dynamics \eqref{eqPNnnnnn1} and its piecewise Chebyshev approximation \eqref{eqPNnn1} on $(t^{w-1},t^{w}]$, it holds that
\begin{equation}\label{eq:err_decay}
\mathscr{E}^w_{\max} \;\le\; C_w\,\alpha_{m,w}^{-M_w},
\end{equation}
where 
\(0<C_w<\frac{2D_w^{(M_w+1)}}{\sqrt{6\pi}}\bigl(\frac{\tau}{4}\bigr)^{M_w+1}\), \(M_w\ge2\), and \(\alpha_{m,w}=\frac{M_w+1}{e}\) with Euler’s number \(e\approx2.718\).
\end{lemma}
\emph{Proof:} The proof is given in the Appendix.

 \begin{theorem} \label{Th:minimum_node}
Let $F^w(t)$ in~\eqref{eqPNnnnnn1} be piecewise continuous on $(t^{w-1}, t^{w}]$, where the window length satisfies $\tau<4$.
Given a desired approximation error $\epsilon_{\mathrm{th}}$ and an initial node count $M_w$, and using the approximation $\hat{F}^w(t)$ in~\eqref{eqPNnn1}, 
the minimum node count that keeps the  $\mathscr{E}^w_{\max}$ within $[\kappa \epsilon_{\mathrm{th}},\, \epsilon_{\mathrm{th}}]$  can be computed as
\begin{equation}\label{eq:node_update}
\bar{M}_{w} =
\begin{cases}
   M_w + \left\lceil \rho_1^w \ln\!\left({\mathscr{E}_{\max}^w}/{\epsilon_{\mathrm{th}}}\right) \right\rceil,\,\,\,\,\,\,\,
    \mathscr{E}_{\max}^w > \epsilon_{\mathrm{th}},\\[6pt]
   M_w, \quad\quad\quad\quad\quad\quad\,\,\,\,\,\,\,
    \kappa \epsilon_{\mathrm{th}} \le \mathscr{E}_{\max}^w \le \epsilon_{\mathrm{th}},\\[6pt]
   M_w + \left\lfloor \rho_2^w \ln\!\left({\mathscr{E}_{\max}^w}/{\kappa \epsilon_{\mathrm{th}}}\right) \right\rfloor, \,
    \mathscr{E}_{\max}^w < \kappa \epsilon_{\mathrm{th}},
\end{cases}
\end{equation}
where $0<\kappa<1$, $\rho_1^w=\frac{1}{\ln(\alpha_{m,w})}$, $\alpha_{m,w}=\frac{M_w+1}{e}$, $\rho_2^w=\frac{1}{\ln(\underline{\alpha}_{m,w})}$, $\underline{\alpha}_{m,w}=\frac{\underline{M}_w+1}{e}$, and $\underline{M}_w$ is the  node count satisfying $\mathscr{E}_{\max}^w>\kappa\epsilon_{\mathrm{th}}$.
\end{theorem}

\noindent \emph{Proof:} The proof is given in the Appendix.

\begin{remark}
Given a tolerance $\epsilon_{\text{th}} \in \mathbb{R}_{>0}$, an error band
$[\kappa\epsilon_{\text{th}},\,\epsilon_{\text{th}}]$ with $0<\kappa<1$ is employed to create a dead zone.
When the error lies within this band, the node count is held constant.
\end{remark}

 \begin{remark}\label{remarrk7}
The node count in \eqref{eq:node_update} uses a conservative error bound $\bar{C}_{w}\,\bar{\alpha}_{m,w}^{-\bar{M}_{w}}$ (see \eqref{eq:err_decay1221} in the Appendix), which is below the threshold $\epsilon_{\mathrm{th}}$. Therefore, \eqref{eq:node_update}  is a sufficient condition only and the actual node count needed to meet $\epsilon_{\mathrm{th}}$ may be less than or equal to the node count $\bar{M}_{w}$.
\end{remark}

%============================================

\section{Adaptive State Estimator And Node Design}\label{sec:state_estimation}
This section presents the design of an adaptive state estimator that leverages approximated system dynamics and a node selection criterion to ensure accurate state estimation.

 \subsection{Adaptive State Estimator Design  } \label{sec:state_estomator}

Unlike traditional ID, the proposed online PS approach approximates the dynamics piecewise as in \eqref{eqPNnn1}. Therefore, the state estimator dynamics can only be  defined piecewise for each window $w$. However, the coefficients $\eta^{w}$ for $w$-th window are computed at the end of the window at time $t^w$ using \eqref{eqPN12} from the data collected between $(t^{w-1},t^{w}]$. Hence, the estimated coefficients are not be available for implementation in the $w$-th window.  Using the available information, i.e., previous window's coefficients $\eta^{w-1}$ and number of Chebyshev nodes $M_{w-1}$, the state estimator dynamics for the $w$-th window can be defined as
\begin{equation}\label{identifier_dynamics1}
    \dot{\hat{x}}(t) = {{\eta }^{{{w-1}^\top}}}{\bar{T}}^{S_{w}}_{M_{w-1}}({t})  - K^{w} \big(x (t^{w-1}) - \hat{x}(t)\big),
\end{equation}
for  $t^{w-1} < t \le t^{w}$, where  \(K^{w} \in \mathbb{R}^{N_p \times N_p}\) represents the gain matrix for  window $w$. Note that ${{\eta }^{{{w-1}^\top}}}{\bar{T}}^{S_{w}}_{M_{w-1}}({t})$ is the dynamics computed using available estimated parameters $\eta^{w-1}$ from $(w-1)$-th window and shifted basis for $w$-th window (super script $S_w$) of order  $M_{w-1}$. 

Although, the state estimator in \eqref{identifier_dynamics1} is implementable, guaranteeing stability for all $t \in \mathbb{R}_{\ge 0}$ may be challenging due to the use of a piecewise approximated dynamics, which may lack sufficient smoothness at the window transitions. In addition, since the shifted Chebyshev polynomials from any window map their domains to $[-1,1]$, the dynamics  $\eta^{w-1}\bar{T}^{S_{w}}_{M_{w-1}}(t)$ in \eqref{identifier_dynamics1} is a delayed version of the approximated dynamics in the previous window ($\eta^{w-1}\bar{T}^{S_{w-1}}_{M_{w-1}}(t)$), as demonstrated in the following proposition, which can further introduce errors in state estimation.

\begin{proposition}\label{prop1_one_window_delay}
\label{thm:delay_chebyshev}
Consider the piecewise approximation of the system dynamics in \eqref{eqPNnn1} on \( (t^{w-2}, t^{w-1}]\), given by  $\hat{F}^{w-1}(t) = \eta^{{w-1}^\top} {\bar{T}}^{S_{w-1}}_{M_{w-1}}({t}),$
and on $(t^{w-1},t^{w}]$ by
$
\hat{F}^{w}(t) = \eta^{{w-1}^\top} {\bar{T}}^{S_{w}}_{M_{w-1}}({t}).$
For equal-width windows $\tau$, it holds that $\hat{F}^{w-1}(t) =\hat{F}^w(t+\tau).$
\end{proposition}

\noindent \emph{Proof:}
The proof is given in the Appendix.

To address these challenges, the continuity of the approximated dynamics and their derivatives must be enforced at the transition between adjacent windows. This requires updating the coefficients $\eta^{w-1}$ corresponding to the $w$-th window.

Let the new approximated dynamics for $w$-the window of the state estimator using $M_{w-1}$ nodes are given by
\begin{equation}\label{ChebApprox2}
\hat{F}^w_\theta(t) = {{\theta }^{{{w}^\top}}}{\bar{T}}^{S_{w}}_{M_{w-1}}({t}), \quad t^{w-1}< t \le t^{w},
\end{equation}
where $ \theta^{w} \in \mathbb{R}^{M_{w-1}+1}$ are the updated coefficients.  For continuity of the dynamics and its derivatives up to order $M_{w-1}$ at $t^{w-1}$, it must hold that
\begin{equation}\label{EQPAR10}
\!\!\!\!\!\left.\frac{d^{p}}{dt^{p}}\big({\eta^{w-1}}^{\top}\bar{T}^{S_{w-1}}_{M_{w-1}}(t)\big)\right|_{t=t^{w-1}}
\!\!\!\!\!\!=\!\!
\left.\frac{d^{p}}{dt^{p}}\big({\theta^{w}}^{\top}\bar{T}^{S_{w}}_{M_{w-1}}(t)\big)\right|_{t=t^{w-1}}\!\!\!\!\!\!\!,
\end{equation}
 for $p=0,\ldots ,{{M}_{w-1}}$. Since the left side of \eqref{EQPAR10} is known, $\theta^{w}$ can be computed by solving the simultaneous equations. 

\begin{lemma}\label{lema:state_estimator_continuity}
Given the continuity condition stated in \eqref{EQPAR10}, the solution $\theta^w$ exists and is unique.
\end{lemma}
\emph{Proof:} The proof is given in the Appendix.

With \eqref{ChebApprox2}, the adaptive state estimator dynamics in \eqref{identifier_dynamics1} on $(t^{w-1},t^{w}]$ can be expressed as
\begin{equation}\label{identifier_dynamics22}
\begin{aligned}
    \dot{\hat{x}}(t)  = {{\theta }^{{{w}^\top}}}{\bar{T}}^{S_{w}}_{M_{w-1}}({t})  - K^{w} (x(&t^{{w-1}}) - \hat{x}(t)), 
    \end{aligned}
\end{equation}
where
\begin{equation}\label{eq1:hat{x}}
     \hat{x}(t) = \begin{cases}
         x(t^{w-1}) & t = t^{w-1}, \\
         \hat{x}(t)  & t^{w-1} < t \le t^{w}.
     \end{cases}
 \end{equation}
At the start of each window, the state estimates in \eqref{identifier_dynamics22} are reset to the measured values via \eqref{eq1:hat{x}}. 

While the using coefficient vector $\theta^w$ enables online state estimation, it inevitably introduces additional approximation error, as demonstrated in the following theorem.

\begin{theorem} \label{th:state-estimator_error_bound}
Consider the dynamics \eqref{Eq: dyanmicform_closed_loop1}, approximated by the Chebyshev expansion \eqref{ChebApprox2}. Let $\theta^{w}$ be computed according to \eqref{EQPAR10}. Then the function approximation error 
$\mathscr{E}^w_\theta(t) \!=\! F^w(t)\! -\! \hat{F}^w_\theta(t)  \text{ for } t\in(t^{w-1},t^{w}]$
is bounded above as
\begin{equation}\label{second approach error modified}
\begin{aligned}
|\mathscr{E}^w_\theta(t)| &\leq \frac{2D_{w-1}^{({M}_{w-1}+1)}}{{{M}_{w-1}}!} \left(\frac{\tau}{4}\right)^{{M}_{w-1}+1}  + L^w \tau  \\ &+ 2\sqrt{{M}_{w-1}+1}\|{\theta}^w\|,
\end{aligned}
\end{equation}
where \( L^w \) is the Lipschitz constant on the \( w \)-th time window.
\end{theorem} 

\noindent \emph{Proof:} The proof is given in the Appendix.

From Theorem \ref{th:state-estimator_error_bound}, the approximation error splits into three parts. The first is the same as the offline Chebyshev bound \eqref{eq:chebyshev_error}. The second and third terms quantify the effect of window length and online parameter estimation errors. To minimize the $\mathscr{E}^w_\theta(t)$ from available data, we define its windowed average and then redefine the node-selection criterion.

 \subsection{Adaptive Node Selection Approach } \label{node_selection}
The error $\mathscr{E}^w_\theta(t)$ is not computable for all $t$, since $F^w(t)$ is unknown.
 With the available information at nodes $t_k^w$,  the average approximation error on $(t^{w-1},t^{w}]$ can be defined as
\begin{equation}\label{h_infty_approx_errorrrr1}
\begin{aligned}\mathscr{E}^w_{\theta,\text{avg}} =\frac{1}{M_w+1} \sum\nolimits_{k=1}^{M_w+1} & \left\| F^w\big(t_k^w\big) - \hat{F}^{w}_\theta\big(t_k^w\big)  \right\|.
 \end{aligned}
 \end{equation}

 \begin{remark}
The approximation error at node points is zero when \(\hat{\mathscr{F}}^{w}(t)\) is obtained from \eqref{eqPNnn1}, i.e., \(\mathscr{F}^w(t_k^w)=\hat{\mathscr{F}}^{w}(t_k^w)\). Since the state estimator utilizes \eqref{ChebApprox2}, the error in \eqref{h_infty_approx_errorrrr1} is nonzero unless the approximation is exact. 
 \end{remark}

  The average approximation error $\mathscr{E}^w_{\theta,\text{avg}}$ in \eqref{h_infty_approx_errorrrr1} can be compared with a desired approximation error  $\epsilon_{\mathrm{th}}$ at each window to dynamically adjust the node count for the next window. Recalling  Theorem \ref{Th:minimum_node}, the adaptive update law for the  node count that keeps the  $\mathscr{E}^w_{\theta,\text{avg}}$ 
within $[\kappa \epsilon_{\mathrm{th}},\, \epsilon_{\mathrm{th}}]$  can be expressed as 
\begin{equation}\label{eq:node_update_law}
    M_{w+1} = \begin{cases}
   M_w +   \left\lfloor \gamma_1^w \ln\left({{\mathscr{E}}^w_{\theta,\text{avg}}}/{{\epsilon}_{\mathrm{th}}}\right) \right\rfloor,   \,\,\,\,\mathscr{E}^w_{\theta,\text{avg}} > \epsilon_{\mathrm{th}},\\
    M_w, \quad\quad\quad\quad\quad\quad\,\,\,\,\,\,\,\,\,\, \kappa \epsilon_{\mathrm{th}} \leq \mathscr{E}^w_{\theta,\text{avg}} \leq \epsilon_{\mathrm{th}}, \\
   M_w + \left\lceil{\gamma}_2^w  \ln\left({{{\mathscr{E}}}^w_{\theta,\text{avg}}}/{\kappa{\epsilon}_{\mathrm{th}}}\right) \right\rceil,   \mathscr{E}^w_{\theta,\text{avg}}\!<\! \kappa\epsilon_{\mathrm{th}},\\
  \end{cases}
\end{equation}
where, to account for the system nonlinearity, parameters $0 < \gamma_1^w, \gamma_2^w < 10$ are set per the proof of Theorem~\ref{Th:minimum_node}.

\begin{remark}
 According to Remark \ref{remarrk7}, \eqref{eq:node_update_law} swaps the rounding operators in \eqref{eq:node_update}. This also prevents round-off errors from adding an unnecessary node. Constants $\rho_1^w,\rho_2^w$ are replaced with adaptive gains $\gamma_1^w,\gamma_2^w$, which provide additional degrees of freedom to account for the rapid change in slope of the nonlinear dynamics. 
\end{remark}

The next lemma and theorem present the boundedness analysis for the parameter and state estimation errors.

%=================================================

\begin{lemma}\label{th_theta_bound_state_estimator}
Consider the dynamics \eqref{Eq: dyanmicform_closed_loop1} on $(t^{w-1},t^{w}]$ and their approximation in \eqref{ChebApprox2}. Let  $\theta^{w}$ is computed using \eqref{EQPAR10}. Then the parameter estimation error $\tilde{\theta}^{w}=\theta^{w^{*}}-\theta^{w}$ is ultimately bounded (UB).
\end{lemma} 
\noindent \emph{Proof:} The proof is given in the Appendix. 

\begin{theorem}\label{th:state_param_error_UUB}
Consider the dynamics \eqref{Eq: dyanmicform_closed_loop1} on \((t^{w-1}, t^{w}]\), approximated by the Chebyshev expansion \eqref{ChebApprox2}, and the adaptive state estimator \eqref{identifier_dynamics22}. Let \(\theta^{w}\) be updated according to \eqref{EQPAR10}. Then the state-estimation error \(\tilde{x}\) and the parameter-estimation error \(\tilde{\theta}^{w}\) are UB, provided that the gain matrix \(K^{w}\) satisfies the Lyapunov equation
\begin{equation}\label{lyap_equation}
Z^w K^w+K^{w^\top} Z^w = -Q^w,
\end{equation}
where $Z^w$, $Q^w$ are symmetric positive definite matrices, provided that $\lambda_{\min}(Q^w)>3$ and $\|\tilde{x}\|^{2} > {\Upsilon}^w/({\lambda_{\min}(Q^w)-3})$, where ${\Upsilon}^w$ is a constant defined in \eqref{eq: Lyapanouv}. 
\end{theorem}

\noindent \emph{Proof:} The proof is given in the Appendix.

%=====================================================
\section{Simulation Results}\label{simulation_results}
This section presents numerical results using the the Stuart--Landau oscillator, given by
\begin{equation}\label{eq:stuart_landau}
\begin{aligned}
\dot{x}_1 &= \big(a - (x_1^2 + x_2^2)\big) x_1 - \omega x_2,\\
\dot{x}_2 &= \big(a - (x_1^2 + x_2^2)\big) x_2 + \omega x_1,
\end{aligned}
\end{equation}
with $a=0.5$ and $\omega=1.5$. The system exhibits a stable limit cycle of radius $\sqrt{a}$.
The simulation parameters were selected as follows: windows length $\tau=0.2\,\text{s}$, desired approximation error $\epsilon_{\text{th}}=10^{-3}$, time horizon $12\,\text{s}$, and sampling period $0.001s$. The initial approximation order was chosen as $M_w=2$ and updated using \eqref{h_infty_approx_errorrrr1} and \eqref{eq:node_update_law} with $\gamma_1^w=0.2$, $\gamma_2^w=0.9$, and $\kappa=0.1$. The system and identifier states were initialized as $[0.5\ 0.5]^\top$ and $[2\ 2]^\top$, respectively, and the identifier was reset in each window using \eqref{eq1:hat{x}}. The gains $K^w$ followed \eqref{lyap_equation} with $Z^w=10I_2$ and $Q^w=\mathrm{diag}(5,4.5)$. Initial coefficients for the first window were chosen as $\eta^{1}=[\,0.05\,\mathbf{1}_{3\times1}\ \ -0.05\,\mathbf{1}_{3\times1}\,]$ where \(\mathbf{1}_{3\times1}=[1 ~~1 ~~1]^\top\).  

\begin{figure}[!ht]
\centering
\includegraphics[width=0.95\columnwidth]{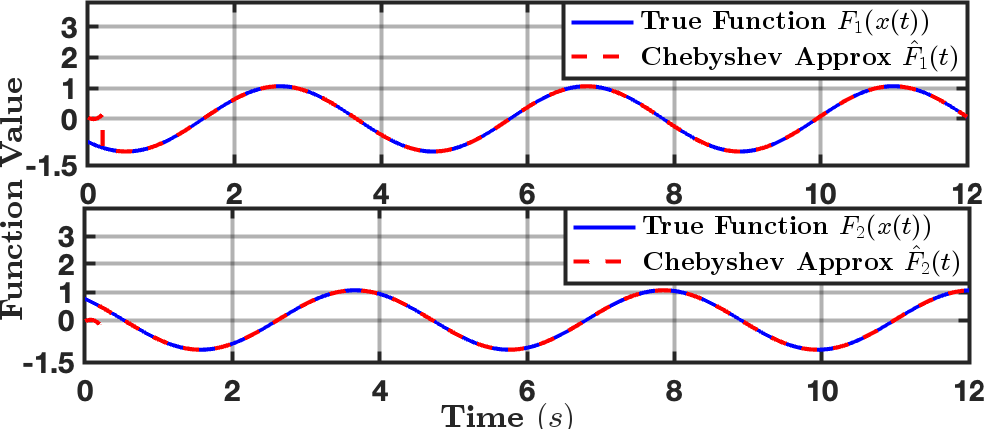}
    \caption{Convergence of the Chebyshev approximation $\hat{F}(t)$ to the true dynamics $F(x(t))$.}
    \label{fig22}
\end{figure}

Fig.~\ref{fig22} shows that the approximated dynamics converge to the true dynamics by $0.4$ s (window $2$). The first window shows a large error due to the initial condition and an insufficient node count (two), which is inadequate to approximate the dynamics. As the node count gets updated in later windows, the error decreases.
\begin{figure}[H]
   \hspace*{-0.2cm}
   \centering
\includegraphics[width=0.95\columnwidth]{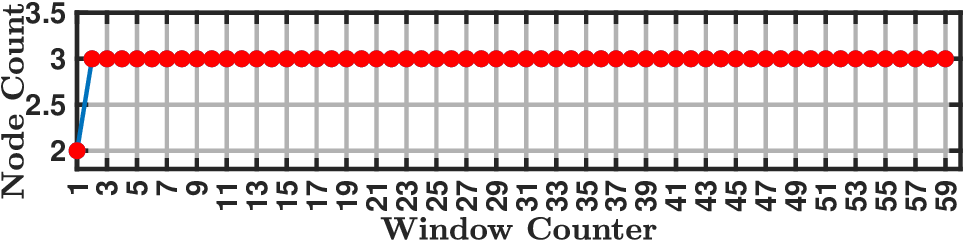}
    \caption{Number of nodes in each time window for approximation.     }
    \label{fig8}
    \vspace{-3mm}
\end{figure}

Fig.~\ref{fig8} shows the adaptive adjustment of node count across moving time windows. Starting from arbitrary initial conditions, the algorithm increases the count to match local complexity, reaching 3 by the second window and driving the error to $10^{-3}$. As the function behavior tends to become nearly linear within each window, the error falls below $10^{-3}$. In concave regions it increases again near $10^{-3}$. With $\kappa=0.1$, since the error remains within $[\kappa \epsilon_{\text{th}}, \epsilon_{\text{th}}]$, the node count stays constant, demonstrating the method’s effectiveness in guaranteeing the desired approximation accuracy. Therefore, compared to neural network–based approximations that rely on periodic sampling (total 12,000 samples), the proposed aperiodic approach offers a significant reduction in sampling frequency (only 35 samples) and, therefore, computational effort.

\begin{figure}[!ht]
   \centering
\includegraphics[width=0.95\columnwidth]{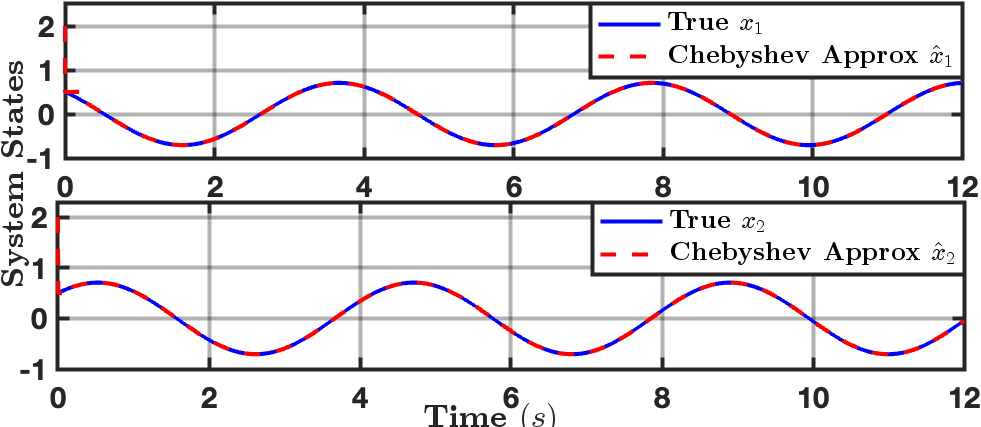}
    \caption{Convergence of Chebyshev approximation  $\hat{x}(t)$ to system state $x$. }
    \label{FigAAr333}
\end{figure}

\begin{figure}[!ht]
   \centering
\includegraphics[width=0.95\columnwidth]{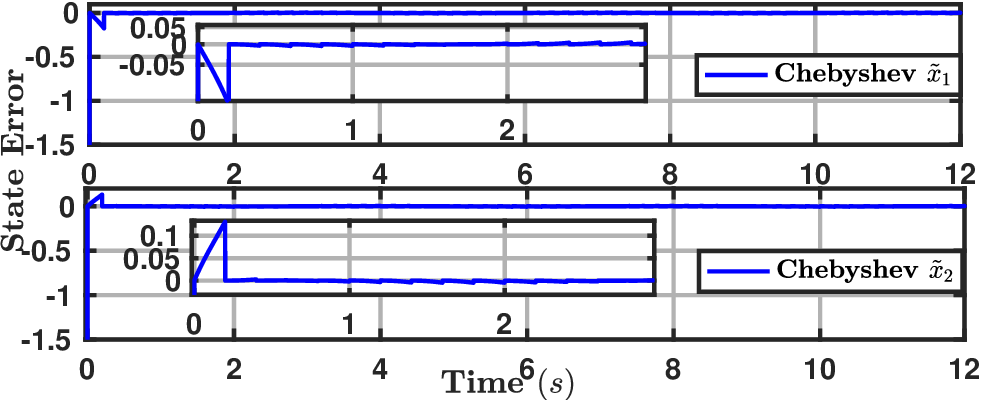}
    \caption{Convergence of the state estimation error $\tilde{x}(t)$.}
    \label{FigAAr33311}
\end{figure}

Figures~\ref{FigAAr333} and \ref{FigAAr33311} depict, respectively, the true state with its estimate $\hat{x}(t)$ and the estimation error $\tilde{x}(t)$. As shown, the estimates converge to the true states within three windows.
The adaptive estimator meets the desired accuracy under aperiodic sampling, which validates the gain design in \eqref{lyap_equation} and the parameter convergence.

%=================================================
\section{Conclusions}\label{sec:conclusion}
This note, presented an online PS method for system ID. The results demonstrate that the proposed online ID framework using first-kind Chebyshev polynomials can accurately approximate nonlinear system dynamics with only a few aperiodic samples per window. The adaptive estimator effectively reconstructed continuous states, and the extended Lyapunov analysis confirmed convergence and boundedness of the parameter and state estimation errors. Numerical studies further validated the analytical findings. This approach  can be employed to design controller by jointly approximating the drift and input vector fields and is included in our future work.

%============================================
% Proofs
%========================================

\numberwithin{equation}{section}
\appendix
\section{Appendix}
%%%%%%%%%%%%%%%%%%%%%%%%%%%%%%%%%%%%%%%%%%%%
% Proof of proposition 1
%%%%%%%%%%%%%%%%%%%%%%%%%%%%%%%%%%%%%%%%%%%%%
\emph{\textbf{Proof of Lemma \ref{lm:exponential_bound_max_error}:}} 
 In \eqref{eq:chebyshev_error}, the factorial term obeys the extended Stirling double inequality \cite{feller1991introduction}, so it holds that
\begin{equation}\label{eq:stirling_bounds}
\sqrt{2\pi M_w}\!\left(\!\frac{M_w}{e}\!\!\right)^{M_w}\!\!\!\!\!\! e^{\frac{1}{12{M_w}+1}} \!\!< \!{M_w}! \!< \!\!\sqrt{2\pi {M_w}}\left(\!\frac{M_w}{e}\!\!\right)^{M_w} \!\!\!\!\!e^{\frac{1}{12{M_w}}}. 
\end{equation}
Taking reciprocals in \eqref{eq:stirling_bounds} for $M_w+1$ yields
\begin{equation}\label{eq:reciprocal}
\frac{1}{(M_w+1)!} \!< \!\!\frac{1}{\sqrt{2\pi (M_w+1)}} \left(\!\frac{e}{M_w+1}\!\right)^{M_w+1}\!\!\!\!\!\!\! e^{-\frac{1}{12(M_w+1)+1}}. 
\end{equation}
From \eqref{eq:reciprocal}, for $M_w\geq2$, the following inequalities hold 
\begin{equation}\label{allineqs}
\begin{aligned}
   \frac{1}{\sqrt{2\pi (M_w+1)}} &\;\le\; \frac{1}{\sqrt{6\pi}},\,\, 
   e^{-\frac{1}{12(M_w+1)+1}} \;<\; 1, \\[6pt]
   \left(\tfrac{e}{M_w+1}\right)^{M_w+1} &\;<\; \alpha^{-(M_w+1)},
     \quad \text{for } \alpha < \tfrac{M_w+1}{e}, \\[6pt]
   \alpha^{-(M_w+1)} &\;<\; \alpha^{-(M_w)},
     \quad \text{for } \alpha > 1.
\end{aligned}
\end{equation}
 From \eqref{eq:chebyshev_error}, with ${\alpha}_{m,w}=\sup\alpha=\frac{M_w+1}{e}$ and using \eqref{allineqs}, the maximum approximation-error is bounded by
\begin{equation}\label{eq:err_decay1}
\mathscr{E}^w_{\max} \leq C_w{\alpha}_{m,w}^{-M_w}, ~t^{w-1} < t \le t^{w}, 
\end{equation}
where $M_w\geq2$ , and $C_w$ satisfying
\begin{equation}\label{cwwww}
    0<C_w<\frac{2D_w^{(M_w+1)}}{\sqrt{6\pi}}(\frac{\tau}{4})^{(M_w+1)}.  
\end{equation}
This completes the proof. \hfill $\blacksquare$

\emph{\textbf{Proof of Theorem 1:}} Based on \eqref{eq:node_update}, the proof is carried out in three  cases. 
\vspace{2mm}

\textbf{Case I:} \textit{For the case \(\mathscr{E}^w_{\max} > \epsilon_{\mathrm{th}}\)}.
To reduce the error below \(\epsilon_{\mathrm{th}}\), an increment \(\Delta M_w = \bar{M}_{{w}} - M_w>0\) in $w$-th window is required such that 
\begin{equation}\label{innnnequallity}
\bar{C}_{w}\,{\alpha}_{{m,w}}^{-(M_{{w}}+\Delta M)} < \epsilon_{\mathrm{th}}, ~t^{w-1} < t \le t^{w},
\end{equation}
where $\bar{M}_w$ is the minimum number of nodes,
and $\bar{C}_w$ satisfies 
\begin{equation}\label{barcc}
0<\bar{C}_w<\frac{2D_w^{(M_w+\Delta M_w+1)}}{\sqrt{6\pi}}\left(\frac{\tau}{4}\right)^{(M_w+\Delta M_w+1)}.
\end{equation}
 Defining $\bar{\alpha}_{m,w}= (\bar{M}_w+1)/{e}$ and since $\bar{M}_w>M_w$ it holds that
\begin{equation}\label{innequallity}
    {\bar\alpha}_{m,w}>{\alpha}_{m,w}.
\end{equation}
From \eqref{innnnequallity} and \eqref{innequallity} it can be verified that
\begin{equation}\label{eq123321}
    \bar{C}_{w}\,{\bar\alpha}_{m,w}^{-\bar{M}_{{w}}}<\bar{C}_{w}\,{\alpha}_{{m,w}}^{-(M_{{w}}+\Delta M)}< \epsilon_{\mathrm{th}}.
\end{equation}
Using \eqref{eq123321} and \eqref{eq:err_decay1}, and denoting $\bar{\mathscr{E}}^w_{\max}$ as the updated maximum error after node increment, leads to
\begin{equation}\label{eq:err_decay1221}
\bar{\mathscr{E}}^w_{\max}\leq \bar{C}_{w}\,\bar{\alpha}_{m,w}^{-\bar{M}_{w}} < \epsilon_{\mathrm{th}}.
\end{equation}
Since \(\mathscr{E}^w_{\max} > \epsilon_{\mathrm{th}}\), by \eqref{eq:err_decay1} and \eqref{eq:err_decay1221} one can get
\begin{equation}\label{innnnequallity2}
\bar{C}_{w}\,\bar{\alpha}_{{m,w}}^{-(M_{{w}}+\Delta M_w)}<{C}_{w}\,{\alpha}_{{m,w}}^{-M_{{w}}} .
\end{equation}
Since ${C}_{w}\,{\alpha}_{{m,w}}^{-M_{{w}}}>{C}_{w}\,{\alpha}_{{m,w}}^{-(M_{{w}}+\Delta M_w)}$, two sub-cases arise.

\textbf{1)} \textit{\textbf{Sub-case I.1}}: For the case \begin{equation}\label{case111}
\bar{C}_{w}\,\bar{\alpha}_{{m,w}}^{-(M_{{w}}+\Delta M_w)}\leq{C}_{w}\,{\alpha}_{{m,w}}^{-(M_{{w}}+\Delta M_w)}.
\end{equation}
Given \eqref{cwwww} and \eqref{barcc}, the condition $\frac{\tau}{4} < 1$ makes $\left(\frac{\tau}{4}\right)^{M_w+1}$ and $\left(\frac{\tau}{4}\right)^{\bar{M}_w+1}$ very small, making $C_w$ and $\bar{C}_w$ tend to zero. So, from \eqref{eq123321} and \eqref{case111} one can write
\begin{equation}\label{case1112}
\bar{C}_{w}\,\bar{\alpha}_{{m,w}}^{-(M_{{w}}+\Delta M_w)}\leq{C}_{w}\,{\alpha}_{{m,w}}^{-(M_{{w}}+\Delta M_w)}<\epsilon_{\mathrm{th}}.
\end{equation}
 Using \eqref{case1112}, \eqref{eq:err_decay1221} satisfies $
\bar{\mathscr{E}}^w_{\max} \leq{C}_{w}\,{\alpha}_{m,w}^{-\bar{M}_{{w}}}< \epsilon_{\mathrm{th}}$. Then taking logarithms both sides yield
\begin{equation}\label{23232112}
\Delta M_w \geq \frac{1}{\ln({{\alpha}}_{m,w})}\left[ \ln\left({{{C}}_w}/{\epsilon_{\mathrm{th}}}\right) - M_w \ln({\alpha}_{m,w}) \right].
\end{equation}
From \eqref{eq:err_decay1}, dividing by $\epsilon_{\mathrm{th}}$ and taking logarithm, it holds 
\begin{equation}\label{eq22212212}
\ln\left({\mathscr{E}^w_{\max}}/{\epsilon_{\mathrm{th}}}\right) \leq \ln\left({{{C}}_w}/{\epsilon_{\mathrm{th}}}\right) - {M}_w \ln({{\alpha}}_{m,w}). 
\end{equation}
Thus, rearranging \eqref{eq22212212} and substituting to \eqref{23232112} leads to
\begin{equation}\label{ineqiuality}
\Delta M_w \geq \frac{1}{\ln({\alpha}_{m,w})}\ln\left({\mathscr{E}^w_{\max}}/{\epsilon_{\mathrm{th}}}\right). 
\end{equation}
Since ${M}_w \geq2$, ${\alpha}_{m,w}\leq\frac{3}{e}$ and $0<\frac{1}{\ln({\alpha}_{m,w})}\leq10.15$. So, denoting $\rho_1^w=\frac{1}{\ln({\alpha}_{m,w})}$, the number of integer nodes can be give as
\begin{equation}\label{eqqqq}
\bar{M}_{w} = M_w + \left\lceil\rho_1^w  \ln\left({\mathscr{E}^w_{\max}}/{\epsilon_{\mathrm{th}}}\right) \right\rceil. 
\end{equation}

\textbf{2)} \textit{\textbf{Sub-case I.2}}: For this sub-case, it holds that
$
\bar{C}_{w}\,\bar{\alpha}_{{m,w}}^{-(M_{{w}}+\Delta M_w)}>{C}_{w}\,{\alpha}_{{m,w}}^{-(M_{{w}}+\Delta M_w)}
$. So, using \eqref{innequallity}, the following inequality holds 
\begin{equation} \label{three_inequality_Case2}
\begin{aligned}
\bar{C}_{w}\,\bar{\alpha}_{{m,w}}^{-(M_{{w}}+\Delta M_w)}  >{C}_{w}\,{\alpha}_{{m,w}}^{-(M_{{w}}+\Delta M_w)} > {C}_{w}\,{\bar\alpha}_{{m,w}}^{-(M_{{w}}+\Delta M_w)}.
\end{aligned}
\end{equation}
From \eqref{three_inequality_Case2}, $\bar{C}_w>C_w$. Hence, \eqref{eq:err_decay1} satisfies
\begin{equation}\label{eq:err_decay12}
\mathscr{E}^w_{\max} \leq \bar{C}_w{{\alpha}}_{m,w}^{-M_w},\,\,\,\,\, ~t^{w-1} < t \le t^{w}. 
\end{equation}
From \eqref{eq123321} and \eqref{eq:err_decay1221},
$\bar{\mathscr{E}}^w_{\max} \leq \bar{C}_{w}\,{\alpha}_{m,w}^{-\bar{M}_{{w}}}< \epsilon_{\mathrm{th}}.$
Taking the logarithm of the right-hand side yields
\begin{equation}\label{eq:err_decay122121}
\Delta M_w \geq \frac{1}{\ln({{\alpha}}_{m,w})}\left[ \ln\left({{{\bar{C}}}_w}/{\epsilon_{\mathrm{th}}}\right) - M_w \ln({\alpha}_{m,w}) \right]. 
\end{equation}
From \eqref{eq:err_decay12}, dividing by $\epsilon_{\mathrm{th}}$ and taking logarithm, we have 
\begin{equation}\label{eq222122121}
\ln\left({\mathscr{E}^w_{\max}}/{\epsilon_{\mathrm{th}}}\right) \leq \ln\left({{{\bar{C}}}_w}/{\epsilon_{\mathrm{th}}}\right) - {M}_w \ln({{{\alpha}}}_{m,w}). 
\end{equation}
From \eqref{eq:err_decay122121} and \eqref{eq222122121}, we can obtain
$\Delta M_w \geq \frac{1}{\ln({\alpha}_{m,w})}\ln\left({\mathscr{E}^w_{\max}}/{\epsilon_{\mathrm{th}}}\right). $
Therefore, the minimum number of nodes is same as in \eqref{eqqqq}. 

\vspace{2mm}
\textbf{Case II:} \textit{For this case $\kappa\epsilon_{\mathrm{th}} \leq \mathscr{E}^w_{\max} \leq \epsilon_{\mathrm{th}}$} .
This case is trivial since the maximum error is within the desired error bound and the number of nodes $M_w$ is the minimum.

\vspace{2mm}
\textbf{Case III}:  \textit{For this case \(\mathscr{E}^w_{\max} < \kappa\epsilon_{\mathrm{th}}\)}.
In this case, nodes are reduced until $\kappa\epsilon_{\mathrm{th}}$. Denote by $\underline{M}_w$ the node count on the last interval with $\mathscr{E}^w_{\max} \geq \kappa \epsilon_{\mathrm{th}}$. When  $\mathscr{E}^w_{\max}  \le \kappa \epsilon_{\mathrm{th}}$, applying \eqref{ineqiuality}, the change in node count satisfies
\begin{equation}\label{eqqqq1}
\Delta{M}_{w} \geq{\rho}_2^w  \ln\left(\kappa{\epsilon}_{\mathrm{th}}/{\mathscr{E}^w_{\max}}\right) , 
\end{equation}
 where ${\rho}_2^w=\frac{1}{\ln({\underline{\alpha}}_{m,w})}$ and $\underline{\alpha}=\frac{\underline{M}_w+1}{e}$.
 In \eqref{eqqqq1}, $\Delta M_w$ is the minimum node-count decrement  to bring $\mathscr{E}^w_{\max}$ to $\kappa\epsilon_{\mathrm{th}}$. So, the minimum number of nodes $\bar{M}_w=M_w-\Delta M_w$ is
\begin{equation}\label{eqqqq21}
\bar{M}_{w} = M_w + \left\lfloor{\rho}_2^w  \ln\left({\mathscr{E}^w_{\max}}/{\kappa{\epsilon}_{\mathrm{th}}}\right) \right\rfloor .
\end{equation}
 This completes the proof. \hfill $\blacksquare$

 \emph{\textbf{Proof of Proposition \ref{prop1_one_window_delay}:}} Define $t$ on $(t^{w-2},t^{w-1}]$ and $\bar t$ on $(t^{w-1},t^{w}]$. With equal interval length $\tau$, we have $t=\bar t-\tau$.
 
From \eqref{eqPNnn1}, the approximation on $(t^{w-1},t^{w}]$ using $\eta^{w-1}$ (estimated on $(t^{w-2},t^{w-1}]$) is
\begin{align}\label{shifted__delay_problem}
\hat{F}^w(\bar{t}) = \eta^{w-1^\top} \bar{T}^{S_{w}}_{M_{w-1}}(\bar{t}), \quad t^{w-1} < \bar{t} \leq t^w.
\end{align}

Using \eqref{vector basisss} and \eqref{shifftedpoly}, on $(t^{w-2},t^{w-1}]$ the basis for window $w-1$ can be expressed as
\begin{equation}\label{shifftt}
\bar{T}_{M_{w-1}}^{S_{w-1}}(t) = \bar{T}_{M_{w-1}}\big(\frac{2t - (t^{w-2} + t^{w-1})}{t^{w-1} - t^{w-2}}\big).
\end{equation}
Similarly, for the interval \((t^{w-1}, t^w]\), i.e., window $w$, the basis
\begin{equation}\label{shifteed for w}
\bar{T}_{M_{w-1}}^{S_{w}}(\bar{t}) = \bar{T}_{M_{w-1}}\big(\frac{2\bar{t} - (t^{w-1} + t^w)}{t^w - t^{w-1}}\big).
\end{equation}
 Using \(\bar{t}=\!t+\tau\), $\tau\!=\!t^{w-1}\!-t^{w-2}$,  $t^{w}\!=\!t^{w-2}\!+2\tau$, \eqref{shifteed for w}  leads  to
\begin{equation}\label{shiftttt}
\bar{T}_{M_{w-1}}^{S_{w}}(\bar{t}) = \bar{T}_{M_{w-1}}\big(\frac{2t - (t^{w-2} + t^{w-1})}{\tau}\big).
\end{equation}
From \eqref{shiftttt} and \eqref{shifftt}, $\bar{T}_{M_{w-1}}^{S_{w}}(\bar{t}) = \bar{T}_{M_{w-1}}^{S_{w-1}}(t)$. Left-multiplying by $(\eta^{w-1})^\top$ and invoking \eqref{shifted__delay_problem} gives
\begin{equation}\label{shifted__delay_problemm}
    \hat{F}^w(\bar{t}) = \hat{F}^w\big(t+\tau\big) = \eta^{w-1^\top} \bar{T}^{S_{w-1}}_{M_{w-1}}(t) = \hat{F}^{w-1}(t).
\end{equation}
This completes the proof. \hfill $\blacksquare$

%======================================
% Proof of Lemma 1
%=======================================
\emph{\textbf{Proof of Lemma \ref{lema:state_estimator_continuity}:}} Stacking \eqref{EQPAR10} for \(p \!=\! 0,...,M_{w-1}\) gives
\begin{equation}\label{matrixx}
\begin{aligned}
&\left.
\begin{bmatrix}
{T}_{0}^{S_{w}}(t) & {T}_{1}^{S_{w}}(t) & \cdots & {T}_{M_{w-1}}^{S_{w}}(t) \\
0 & \frac{d{T}_{1}^{S_{w}}(t)}{dt} & \cdots & \frac{d{T}_{M_{w-1}}^{S_{w}}(t)}{dt} \\
\vdots & \vdots & \ddots & \vdots \\
0 & 0 & 0 & \frac{d^{M_{w-1}}{T}_{M_{w-1}}^{S_{w}}(t)}{dt^{M_{w-1}}}
\end{bmatrix}
\right|_{t = t^{w-1}}\!\!\!\!\!\!\!\!\!\!
\begin{bmatrix}
\theta_{0j}^{w} \\
\theta_{1j}^{w} \\
\vdots \\
\theta_{M_{w-1}j}^{w}
\end{bmatrix} \\[10pt]
&=
\left.
\begin{bmatrix}
\eta_{j}^{w-1^{\top}}\,\bar{T}_{M_{w-1}}^{S_{w-1}}(t) \\
\eta_{j}^{w-1^{\top}}\,\frac{d\bar{T}_{M_{w-1}}^{S_{w-1}}(t)}{dt} \\
\vdots \\
\eta_{j}^{w-1^{\top}}\,\frac{d^{M_{w-1}}\bar{T}_{M_{w-1}}^{S_{w-1}}(t)}{dt^{M_{w-1}}}
\end{bmatrix}
\right|_{t = t^{w-1}}\!\!\!\!\!\!\!\!\!\!,\quad j = 1, 2, \dots, N_{\mathscr{P}}.
\end{aligned}
\end{equation}
With \eqref{shifftedpoly} and $T_i^{S_w}(t^{w-1})\!\!=\!T_i(-1)$, each derivative in \eqref{matrixx} can be expressed as
\begin{equation}\label{eq12332121}
\begin{aligned}
\left.\frac{d^{p}T_{i}^{S_{w}}(t)}{dt^{p}}\right|_{t=t^{w-1}}=\left(\frac{2}{\tau}\right)^{p}
    \left.\frac{d^{p}T_{i}(t)}{dt^{p}}\right|_{t=-1}\!\!\!.
\end{aligned}
\end{equation}

By \cite[Eq.~(3.35)]{johnson1996chebyshev}, the $p$-th derivative of $T_i(t)$ at $t=-1$ admits a non-zero closed-form expression. So from \eqref{eq12332121}, it holds
$\left.\frac{d^{p}T_{i}^{S_{w}}(t)}{dt^{p}}\right|_{t=t^{w-1}}\neq0$. Hence the upper triangular matrix in \eqref{matrixx} has a nonzero diagonal and is invertible, so a unique $\theta^{w}$ exists. This completes the proof. \hfill $\blacksquare$

%========================================
%Proof of Tehorem 2
%===================================
 \emph{\textbf{Proof of Theorem \ref{th:state-estimator_error_bound}:}}
Recalling \eqref{ChebApprox2}, the function approximation error can be written as
\begin{align}\label{eqP114031006}
        &\mathscr{E}^w_\theta(t) \!=\! F^w(t)\! -\! \hat{F}^w_\theta(t) 
        \!=\! \big(F^w(t) - F^w({t}^{w-1})\big) 
    \!+\!\big(F^w({t}^{w-1}) \nonumber\\  & -\hat{F}_\theta^{w}({t}^{w-1})\big) \!+\! \big(\hat{F}_\theta^{w}({t}^{w-1}) \!-\! \hat{F}^w_\theta(t)\big), {t}^{w-1} \!\!< t \le {t}^{w}\!\!.
    \end{align}
With \( |t - t^{w-1}| \!\!<\!\! \tau \) on $(t^{w-1},t^{w}]$, the first term in \eqref{eqP114031006} satisfies
\begin{equation}\label{eq99}
\|F^w(t) \!- \!F^w(t^{w-1})\|\! \leq\! L^w |t - t^{w-1}| \!\leq\! L^w \tau.
\end{equation}

From \eqref{EQPAR10} for $p=0$, we have $\hat{F}^{w-1}(t^{w-1})=\hat{F}_\theta^w(t^{w-1})$, and trivially ${F}^{w-1}(t^{w-1})={F}^w(t^{w-1})$. Thus, using \eqref{eq:chebyshev_error}, the second term in \eqref{eqP114031006} is given by
\begin{equation}\label{eq100}
\|\big(F^w({t}^{w-1}) \!- \hat{F}_\theta^{w}({t}^{w-1})\big)\|\!
\;\!\!\le\;\!\!\!
\frac{2D_{w-1}^{({M}_{w-1}+1)}}{{M}_{w-1}!}\!
\left(\frac{\tau}{4}\right)^{{M}_{w-1}+1}\!\!\!\!\!\!\!\!\!\!\!\!\!,
\end{equation}
and
$
D_{w-1}^{({M}_{w-1}+1)} \;=\; 
\max_{\xi \,\in\, [t^{w-2},\,t^{w-1}]}
\bigl\|F^{({M}_{w-1}+1)}(\xi)\bigr\|
$
is a bounded scalar.

Finally, using \eqref{ChebApprox2}, the third term in \eqref{eqP114031006} is given by
\begin{align}\label{eqq100}
\hat{F}_\theta^{w}(t^{w-1}) - \hat{F}_\theta^{w}(t) &= 
\theta^{w^\top} \bar{T}^{S_w}_{{M}_{w-1}}\bigl(t^{w-1}\bigr)  - \theta^{w^\top} \bar{T}^{S_w}_{{M}_{w-1}}(t).
\end{align}
Recall that $
\|\bar{T}^{S_w}_{M_{w-1}}(t^{w-1})\| = \sqrt{M_{w-1}+1}
$ and $
\|\bar{T}^{S_w}_{M_{w-1}}(t)\| \;\le\; \sqrt{M_{w-1}+1}
$. Using the inequality \(\|A - B\| \le \|A\| + \|B\|\), \eqref{eqq100} can be written as
\begin{equation}\label{eq1012}
\begin{aligned}
&\|\hat{F}_\theta^{w-1}(t^{w-1}) - \hat{F}_\theta^{w-1}(t)\| \leq\! \|\theta^w\|
\Bigl( \|\bar{T}^{S_w}_{{M}_{w-1}}(t^{w-1})\| 
\! \\ & +\! \|\bar{T}^{S_w}_{{M}_{w-1}}(t)\| \Bigl) \leq 2\,\!\sqrt{{M}_{w-1}\!+\!1}\,\|{\theta}^w\|.
\end{aligned}
\end{equation}
By combining \eqref{eq99}, \eqref{eq100}, and \eqref{eq1012}, the error bound in \eqref{second approach error modified} can be reached.
This completes the proof.\hfill $\blacksquare$

%==============================================
%Proof of Lema 2
%===============================================
\emph{\textbf{Proof of Lemma \ref{th_theta_bound_state_estimator}:}}
By the Chebyshev property, all entries of $\bar{T}^{S_w}_{M_w}(t)$ lie in $[-1,1]$ for $t\in(t^{w-1},t^{w}]$. Hence, \eqref{EQPAR10} gives
\begin{equation}\label{eqPN113}
    R^w_0\le \bar{\mathbb{T}}_{{{M}_{w}}}^{{{S}_{w}}}\left( t \right)\bar{\mathbb{T}}_{{{M}_{w}}}^{{{S}_{w}}^{\top}}\left( t \right)+R^w_0<{{\left( {{M}_{w}}+1 \right)}^{2}}{{I}_{{({M}_{w}}+1)}}+R^w_0.
\end{equation}
It can be verified from \eqref{eqPN113} that ${{\left( \bar{\mathbb{T}}_{{{M}_{w}}}^{{{S}_{w}}}\left( t \right)\bar{\mathbb{T}}_{{{M}_{w}}}^{{{S}_{w}}^{\top}}\left( t \right)+R^w_0 \right)}^{-1}}\leq {{\left( R^w_0 \right)}^{-1}}.$ 
Thus, the inverse in \eqref{eqPN12} is bounded. From stability of \eqref{Eq: dyanmicform_closed_loop1}, $\dot{{X}}^{\,w}(t)$ is bounded. Hence $\eta^{w-1}$ from \eqref{eqPN12} and $\theta^{w}$ from \eqref{EQPAR10} are bounded. Since the actual coefficients $\theta^{{w}^*}$ is bounded. Therefore $\tilde{\theta}^{\,w} = \theta^{{w}^*} - \theta^{w}$ is also bounded. This completes the proof. \hfill $\blacksquare$

%======================================
% Proof of Theorem 3
%=======================================

 \emph{\textbf{Proof of Theorem \ref{th:state_param_error_UUB}:}}
The state measurement error due to aperiodic availability of the state at the identifier can be defined as
$e_s^w=x^{w-1}-{x}$.
Since the system states are bounded, $e^w_s$ is also bounded, i.e., $\|e_s^w\| < e_{s,M_w} $.  It holds that
\begin{equation}\label{eq:tilde-x-second-kind}
\begin{aligned}
&\dot{\tilde{x}}(t)
= \tilde{\theta}^{w^{\top}}\bar{T}_{M_{w-1}}^{S_w}(t)
   + K^{w}\big(x(t^{w-1}) - \hat{x}\big) + \bar\epsilon^{\,w} \\
&= \tilde{\theta}^{w^{\top}}\bar{T}_{M_{w-1}}^{S_w}(t)
   \!+\! K^{w}(e_s^{w} + \tilde{x}) \!+\! \bar\epsilon^{\,w}\!\!,\, t^{w-1} \!\!< t \le t^{w}\!.
\end{aligned}
\end{equation}
Consider the Lyapunov function candidate over $t^{w-1} < t \le t^w$ as $\mathscr{V}^w= \tilde{x}^{^\top} Z^w \tilde{x}\!+ \frac{1}{2} \tilde{\theta}^{{w} ^\top}  \tilde{\theta}^{w}.$
 
 Recall that $\theta^{w}$ and $\hat{\theta}^{w}$ are constant on $(t^{w-1},t^{w}]$, hence $\tilde{\theta}^{w}$ is constant and $\dot{\tilde{\theta}}^{w}=0$. Thus, the Lyapunov first derivative is 
\begin{align}\label{eq: Lyapanouv process}
\dot{\mathscr{V}}^w &=   \text{Sym}(\tilde{x}^\top Z^w \tilde{\theta}^{{w}^\top} \bar{T}_{M_{w-1}}^{S_w}(t))+\text{Sym}(\tilde{x}^\top Z^w K^w e_{s}^w) \nonumber \\
& +\text{Sym}(\tilde{x}^\top Z^w {\bar\epsilon}^w)+\tilde{x}^\top\big(Z^w K^w+K^{w^\top} Z^w\big) \tilde{x}. 
\end{align}
By Young’s inequality, $\text{Sym}( {{X}^{\top}}Y )\le {{X}^{\top}}X+{{Y}^{\top}}Y$; hence
$$
\text{Sym} \big(\tilde{x}^\top Z^w \tilde{\theta}^{{w}^\top} \bar{T}_{M_{w-1}}^{S_w}(t)\big) \leq 
\|\tilde{x}\|^{2}+\big\|Z^w \tilde{\theta}^{{w}^\top} \bar{T}_{M_{w-1}}^{S_w}(t)\big\|^{2},
$$ $
   \text{Sym}(\tilde{x}^\top Z^w K^w e_{s}^w)\leq \|\tilde{x}\|^{2}+\left\|Z^w K^w e_{s}^w\right\|^{2},
$ 
and
$
\text{Sym}(\tilde{x}^\top Z^w \epsilon^w(t))  \leq \|\tilde{x}\|^{2}+\|Z^w {\bar\epsilon}^w\|^{2}
$.
 Recall $\|\bar\epsilon
^w\| \le \epsilon_M^w, \|\bar{T}^{S_w}_{M_{w-1}}(t)\| \le 1$ and by Lemma~\ref{th_theta_bound_state_estimator}, $\|\tilde{\theta}^{w}\| \le \epsilon^w_\eta$. 
So with \eqref{lyap_equation}, the Lyapunov first derivative is bounded by
\begin{align}\label{eq: Lyapanouv}
\dot{\mathscr{V}}^w \leq & -\left( \lambda_{\min }(Q^w) - 3\right)\|\tilde{x} \|^2+ {\Upsilon}^w, 
\end{align}
where  ${\Upsilon}^w = \|Z^w\epsilon^w_{{\eta}} \|^2  + \|Z^w{\epsilon}_{M}^w\|^2 +  \|Z^w K^w e_s^w\|^2$. From \eqref{eq: Lyapanouv}, $\dot{\mathscr{V}}^w<0$ as long as $
  \lambda_{\min }(Q^w)>3$ and $    \|\tilde{x}\|^{2}>\frac{{\Upsilon}^w}{\lambda_{\min }(Q^w) - 3}$. Therefore, $\tilde{x}$ is UB
with the bound 
$    \|\tilde{x}\|^{2} \le \frac{{\Upsilon}^w}{\lambda_{\min }(Q^w) - 3}$. Further, since the estimated state is reset to the true state at each window transition, we have $\tilde{x}(t^w)=0$ at all window transition times, and it is bounded. Therefore, it follows that $\tilde{x}(t)$ is UB.
This completes the proof. \hfill $\blacksquare$

%===========================
%======================================
% References
%=======================================
%=====================================
\bibliographystyle{IEEEtran}  
\bibliography{References/references}

%===================================================

\end{document}